\documentclass[conference]{IEEEtran}
\usepackage{graphicx}
\usepackage[cmex10]{amsmath}
\newcommand{\vy}{\vec{y}}
\newcommand{\vtau}{{\vec{\tau}}}
\newcommand{\vb}{\vec{b}}

\newcommand{\vomega}{\vec{\omega}}
\newcommand{\vnu}{\vec{\nu}}
\newcommand{\ms}{{\bf s}}
\newcommand{\mx}{{\bf x}}
\newcommand{\sk}{{\vec{s}_k}}

\newcommand{\<}{\left\langle}
\renewcommand{\>}{\right\rangle}
\newcommand{\mZ}{{\mathcal Z}}
\newcommand{\mQ}{{\mathcal Q}}

\begin{document}
\title{Randomness and metastability in CDMA paradigms}

\author{\IEEEauthorblockN{Jack Raymond , David Saad}
\IEEEauthorblockA{Aston University, Neural Computing Research
Group,
Birmingham, B4 7ET, UK \\
Email: raymonjr@aston.ac.uk}
}

\maketitle

\begin{abstract}
Code Division Multiple Access (CDMA) in which the signature code
assignment to users contains a random element has recently become
a cornerstone of CDMA research. The random element in the
construction is particularly attractive in that it provides
robustness and flexibility in application, whilst not making
significant sacrifices in terms of multiuser efficiency. We
present results for sparse random codes of two types, with and
without modulation. Simple microscopic consideration on system
samples would suggest differences in the phase space of the
two models, but we demonstrate that the thermodynamic results and
metastable states are equivalent in the minimum bit error rate detector. We 
analyse marginal properties of interactions and also make analogies to constraint satisfiability problems
in order to understand qualitative features of the decoding and metastable states.
This may have consequences for developing algorithmic methods 
to escape metastable states, thus improving decoding performance.
\end{abstract}
\IEEEpeerreviewmaketitle
\section{Introduction}
The area of multiuser communications is one of great interest from
both theoretical and engineering perspectives~\cite{Verdu:MD}.
Code Division Multiple Access (CDMA) is a particular method for
allowing multiple users to access channel resources in an
efficient and robust manner, and plays an important role in the
current standards for allocating channel resources in
wireless communications. CDMA utilises channel resources highly
efficiently by allowing many users to transmit on much of the
bandwidth simultaneously, each transmission being encoded with a
user specific signature code. Disentangling the information in the
channel is possible by using the properties of these codes and
much of the focus in CDMA research is on developing efficient
codes and decoding methods.

A typical CDMA
paradigm is that bandwidth is broken into $N$ discrete Time-Frequency blocks (chips)
with each of $K$ users being assigned a user code ($\sk$) known by the base station, the set of all user codes being $\ms$ (the code). The user code
gives the amplitude and phase by which to modulate transmission of the scalar symbol on each chip.
The signal ($\vy$) received on $N$ chips by the base station is then an interfering (additive) combination of the users' modulated symbols
corrupted during transmission by a fading factor $F_{k \mu}$ and some signal noise ($\nu_\mu$). Assuming perfect synchronisation of the chips the symbols received on each chip are independent and given by
\begin{equation}
  y_\mu = \nu_\mu + \sum_{k=1}^K b_k F_{k \mu} s_{k\mu}\;. \label{channel}
\end{equation}
We focus on a standard channel type (BIAWGN): the Additive White
Gaussian Noise channel (AWGN), employing Binary Phase Shift Keying
(BPSK). The following parameterisations are assumed: the scalar
symbol sent by user $k$ is a bit $b_k=\pm 1$ with probability
$P_{b_k}(b)=\frac{1}{2}$; the noise is Gaussian with zero mean and
variance $\sigma_0^2$ for all chips; prefect power control applies
so that the fading factor $F_{k \mu}=1$; each code element $s_{\mu
k}=\pm A$, where $A$ is the amplitude of the transmission by user
$k$ on chip $\mu$. Generalisations of the model most often
consider the requirement for perfect synchronisation and power
control. Real CDMA applications also have to deal with
idiosynchracies in hardware and environmental conditions not easy
to treat in a generalised analysis, this has not prevented its
updake in some modern wireless communication standards.


This paper follows previous theoretical analyses (e.g.
~\cite{Tanaka:SMA,Yoshida:ASS,Montanari:BPB,Raymond:SS}) in
studying codes which are randomly generated for each system from
some ensemble. The canonical random CDMA ensemble is the dense one
in which all chips are transmitted upon~\cite{Verdu:MD}. In the
sparse ensemble we consider here (\ref{sp}) only a small number of
chips $O(C)$ are accessed by each user, a less studied system.
However there are a number of reasons why the sparse ensemble
first examined in ~\cite{Yoshida:ASS} may be more practical, based
on its closer similarity to FH/TH-CDMA and the ability to apply
fast message passing algorithms in decoding. In addition, one can
converge towards the properties of the dense ensemble by
increasing the mean user connectivity $C$ only moderately. It has
been shown, for a sparse connectivity model in which the mean user
connectivity is large but much smaller than $K$, that the
properties become indistinguishable from the dense channel in
cases where BP converges~\cite{Guo:MDSS}.

The sparse codes consist of a sparse connectivity matrix and a
modulation part sampled according to
\begin{eqnarray}
  P_\ms(\mx) &\propto& \prod_k \prod_\mu\left[\left(1-\frac{L}{K}\right)\delta_{x_{\mu k}} + \frac{L}{K} \phi(x_{\mu k}) \right]\label{sp}\\
  \phi(x) &=& \frac{1}{2}(\delta_{x,A} + \delta_{x,-A}) \label{phi}\;.
\end{eqnarray}
The modulation of non-zero elements in the codes is described by $\phi$ which can be BPSK (as shown) or unmodulated $\phi(x)=\delta_{x,A}$, with
the amplitude of transmission ($A=1/\sqrt{L}$) chosen for normalisation purposes so that the Power Spectral density $\mQ$, a representative measure of signal to noise ratio, may be taken as $1/(2 \sigma_0^2)$. The mean chip and user connectivities are $L$ and $C$, respectively, such that the load $\alpha\!=\!L/C\!=\!K/N$.

Two problems with the basic sparse ensemble (\ref{sp}) at low connectivity is significant asymmetry in bandwidth access for users, with a fraction of users being entirely disconnected. Analogously the utilisation of chips will not be uniform, with some chips unutilised. These problems can be overcome by enforcing regularity of the following forms:
\begin{eqnarray}
 P_\ms(\mx) &\propto&  \prod_k \left[\delta\left(\sum_\mu^N (1 - \delta_{x_{\mu k}}) - C\right)\right] \label{userreg}\;,\\
&\propto&  \prod_k \left[..\right] \prod_\mu  \left[\delta\left(\sum_k^K (1 - \delta_{x_{\mu k}}) - L\right)\right]\;, \label{chipreg}
\end{eqnarray}
in addition to modulation though $\phi$. It turns out that
constraining users to access exactly $C$ chips (\ref{userreg}) is
very important in attaining near optimal performance for high
$\mQ$, whereas enforcing, in addition, chip regular access
(\ref{chipreg}) produces only marginally improved
performance~\cite{Raymond:SS} and may be difficult to implement in
practice. In this paper we consider ensembles with both chip and
user regular constraints (\ref{chipreg}) throughout since it makes
certain aspects of the analysis simpler; we anticipate results to
be qualitatively similar with only the user-regular constraint
(\ref{userreg}).

The theoretical information capacity, and theory of Bayes optimal decoding requires knowledge of the likelihood of transmitted bits
\begin{equation}
  P_{\vy | \vb}(\vtau) \propto \int  \prod_\mu \left[ \delta\left(y_\mu - \sum_k s_{\mu k} \tau_k + \omega_\mu\right)\right] {\hat P}_{\vnu}(\vomega) d\vomega
\end{equation}
where ${\hat P}_{\vnu}$ is the assumed chip noise distribution to be marginalised over. If one considers a Gaussian channel noise model, of variance $(\sigma_0)^2/\beta$ (i.e assumption possibly incorrect by a factor $\beta$), then the righthand side is simplified
\begin{equation}
  P_{\vb | \vy}(\vtau) \propto \prod_\mu \exp\left\{-\beta \mQ \left(y_\mu - \sum_k s_{\mu k} \tau_k\right)^2\right\}\;.
\end{equation}
Statistical physics provides a concise framework to analyse this quantity. First we define a Hamiltonian by connection with the likelihood
\begin{equation}
{\cal H}(\vtau) = \mQ\sum_\mu \left( \nu_\mu + \sum_k s_{\mu k} (b_k - \tau_k) \right)^2\;, \label{Ham}
\end{equation}
where $y_\mu$ is written in terms of its constituent components (\ref{channel}) and $\tau_k$ is a candidate value of the sent bit. From this one can construct the self-averaging free energy.
\begin{equation}
  f = \< -\frac{1}{\beta N} \log \sum_\vtau \exp \{-\beta {\cal H}(\vtau)\} \>\;.\label{fd}
\end{equation}
The average $\<\>$ denotes throughout the paper an average over $\vy$ and codes $\ms$ sampled according to the appropriate ensemble.
The motivation for studying the self-averaged free energy that this is a generating function for many interesting statistics attainable by decoders, averaged over samples of the system. It can be observed that for CDMA the performance measures, such as bit error rate and spectral efficiency, are self-averaging -- rapidly converging to some fixed values as the number of users increase. The bit error rate is mean overlap of the sent and decoded bits $\frac{1}{K}\langle(\vb . \vtau)\rangle$, the spectral efficiency is the mutual information between the sent bits and the received signal $I(\vb,\vy)$ and is affine to the free energy. By taking the limit $K\rightarrow\infty$ we are able to attain an exact description for these fixed points, thereby providing a good indication of performance.
We assume throughout this proceedings that $\beta=1$, analysis of the free energy thereby corresponds to the performance of a detector which minimises the bit error rate.
%
\subsection{Overview of results for BPSK} \label{prevresults}
%
\begin{figure}[!t]
\centering
\includegraphics[width=2.5in]{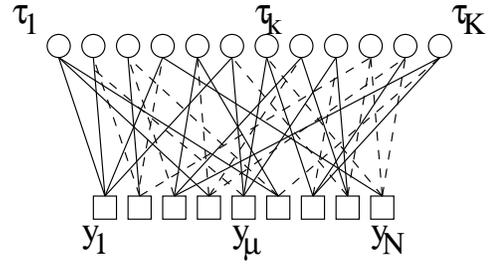}
\caption{\label{graphicalmodels} The inference problem can be represented by a graphical model: a Tanner (or factor) graph. Each factor (square) represents an interaction and each bit (circle) denotes a dynamical variable $\tau_k$ which is to be optimised given the topology and observable values. The observables in this case are the signal $y_\mu$ associated to each node, and the code $\ms$--(dashed/solid lines can be used to indicate  modulation by $\pm A$ in components $s_{\mu k}$). Above is a representation for a small sparse regular graph (\ref{chipreg},\ref{userreg}) with $L\!=\!4$ $C\!=\!3$.
}
\end{figure}
\begin{figure}[!t]
\centering
\includegraphics[width=2.5in]{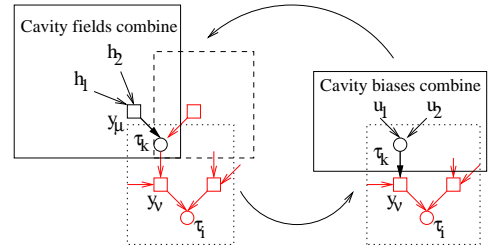}
\caption{\label{popdyns} The fixed points of the self consistent equations are in quantities $h$ and $u$ which have an interpretation in terms of messages passed on (sub)graphs of the graphical model (\ref{graphicalmodels}). If one knows the log likelihood ratio $u_{\mu k}$ of bit $b_k$ given only one of its neighbours $\mu$, then assuming these likelihoods to be independent (as is valid on a tree), one can construct the conditional likelihood of $b_k$ given all its neighbours excluding $\nu$ (or log likelihood ratio $h_{k \nu}$). One can then use $h_{k \nu}$ to construct log likelihoods ($u_{\nu i}$) for subsequent variables in the tree. By such a process, the distribution of $\{h\}$ and $\{u\}$ may converge at sufficient depth in the tree to values independent of the inputs -- such a solution is a viable solution to a population dynamics algorithm. The convergence properties and stability of solutions is closely related to standard decoding algorithms: the sum product algorithm or belief propagation. \vspace*{-4mm}}
\end{figure}
\begin{figure}[!t]
\centering
\includegraphics[width=2.5in]{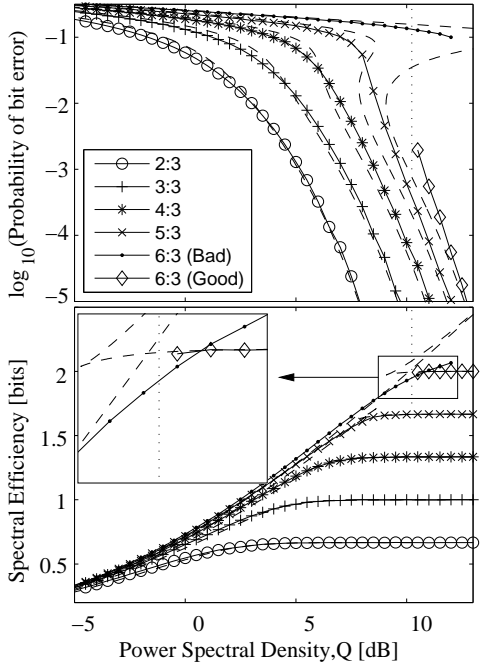}
\caption{\label{fren} The figures show the spectral efficiency (affine to the free energy)
and bit error rate  for a number of cases of $\alpha$ as
indicated by $K\!\!:\!\!N$. The solid curves represent locally
stable solutions of the population dynamics procedure for a sparse
ensemble, dashed curves show the exact results for
the $\mQ$-equivalent densely spread CDMA system -- the curves are qualitatively similar in both quantities, except in the existence of one additional (unstable) solution in the dense case (middle curve). The similarity extends to the metastable ranges, we consider the sparse ensemble results in detail. 
The sparse ensemble
is fully regular with $C\!=\!3$ and $L\!=\!2,..,6$ in agreement
with the ratio $\alpha$. For small loads $\alpha$ a unique
solution is found in both cases, which is the valid thermodynamic
(information theoretic) solution. For the sparse case at
sufficiently large $\alpha$ (case 6:3) the solution becomes
multivalued. 
Lower figure: The
thermodynamic solution is the curve of lowest
spectral efficiency, the other being metastable; there is a second
order transition between the two solution with increasing
$\mQ$. The
inset shows in detail the region in which the dense and sparse
codes undergo thermodynamic second order transitions with $\alpha=2$.
Upper figure: This demonstrates the bit error rate for comparable parameterisations. This figure indicates a large performance gap between the two locally stable solutions in the metastable regime: a bad and good solution exist in terms of decoding. The
vertical dashed line indicates the smallest $\mQ$ at which metastability occurs in the sparse code for the 6:3 case: beyond this point in the metastable regime the bad solution performance is typically attained by belief propagation even if this is only a metastable solution.
\vspace*{-4mm}}
\end{figure}
%

For sparse ensembles with BPSK the equilibrium and dynamical properties are similar to the dense case~\cite{Tanaka:SMA}, becoming more so as $L$ increases~\cite{Guo:MDSS}.
If one calculates the free energy of the sparse ensemble by the cavity or replica method~\cite{Mezard:SGT} one attains under assumptions of a single pure state a site factorised expression for the free energy, determined by the solution to a set of self consistent field and bias distributions (saddlepoint equations) ~\cite{Raymond:SS}. These results are presented for later comparison (\ref{hatW_nomod})
\begin{eqnarray}
  W(h)\! \!&\propto&\!\!  \int \prod_{c=1}^{C-1} \left[ du_c {\hat W}(u_c)\right] \delta\left(h - \sum_{c=1}^{C-1} u_c\right) \nonumber\\
  {\hat W}(u) \!\!&\propto&\!\!  \int \prod_{l=1}^{L-1} \left[ W(h_l) dh_l\right] \prod_{l=1}^{L} \left[\phi(x_l) dx_l\right] P_\nu(\omega)d\omega \nonumber\\
\!\!&\times&\!\! \delta\left(u - \sum_{\tau_L} \tau_L \log(\mZ(\tau_L))\right) \; \label{saddle}\\
  \mZ(\tau_L) \!\!&=&\!\!\sum_{\vtau} \!\exp\left\{\! -\! \mQ \left(\!\omega +\!\sum_{l=1}^{L} \!x_l(1-\tau_l)\right)^2 + \!\sum_l h_l \tau_l\right\} \; \nonumber
\end{eqnarray}
where $P_{\nu}$ is the true chip noise probability distribution.
The distributions are over a set of {\it cavity biases} $u$ and
{cavity fields} $h$. These variables may be interpreted within a
graphical framework of the inference problem
(Fig.~\ref{graphicalmodels}), as log-likelihood (of correct
decoding) ratios in two types of sub-graphs (Fig.~\ref{popdyns}).
From these distributions one can calculate the free energy, bit
error rate and other properties. The equations may be solved
numerically by population dynamics~\cite{Raymond:SS}, which is
implemented as a late	 propagation (decoding) algorithm on a
tree. This processes allows a numerical determination of the free
energy and tests of ergodicity breaking. We find a unique
thermodynamic solution at all $\mQ$, but also a significant
metastable solution for a range of parameters (Fig.~\ref{fren}).

We may distinguish the metastable states in this range of parameters as bad and good (higher or lower bit error rate).
The population dynamics algorithm tends to find the bad solution
from most initial conditions, only those initial conditions which
are of very low bit error rate (a set of cavity biases strongly
correlated with $\vb$) appear to converge towards the good
solution. It appears the bad solution is easy to reach by
implementation of population dynamics regardless of whether it is
the thermodynamically dominant state. This is interesting since
population dynamics appears to mirror the behaviour of many
decoding algorithms on even relatively small systems, which
struggle to achieve good bit error rates in this region. In the
real decoding problem one does not begin the decoding already with
a good estimate of $\vb$, and so one may be stuck with a
suboptimal estimate even where a much better estimate may be found
(in principle) for almost all decodings.


In both the dense and sparse cases there is a unique
thermodynamically stable state. One can hope to achieve the
information capacity of the thermodynamic state by clever
algorihms based on some global insight. The problem is that local
search based optimisation appears insufficient. In the case of no
metastability, local search methods attain the optimal
solution~\cite{Guo:MDSS,Raymond:SS} with various principled
modifications suggested~\cite{Kabashima:MDA}. In the case of
metastability one might apply a principle of guesswork combined
with BP to allow efficient searching of the space. Such a
method~\cite{Measson:MC} has been demonstrated for certain types
of channel, unfortunately not so far the BIAWGN we consider. In
the following sections we consider how the similarity between the
phenomena in dense and sparse systems, combined with a
consideration of marginal interaction distributions, might
characterise the bad metastable solution and how such insight
might be used to supplement local search methods.

\subsection{A sparse model without modulation}
As a way to further understand the microscopic basis of metastability we propose the following model to investigate the sparse ensemble for the case of no modulation, $\phi(x)\!=\!\delta_{x,A}$.
Unlike the dense model, the disorder in the connectivity structure is sufficient to recover information even without modulation. Given that the graphical structure is identical to the modulated sparse ensemble, decoding may be achieved by similar methods (belief propagation based local search).

Working with either the cavity or replica methods one can attain a site factorised set of functional relations analogous to (\ref{saddle}). In the former case we had two distributions containing information on the probabilty of correct bit reconstruction (on two types of subgraph). In the unmodulated case we replace each of these distributions by two, because the probability of correct bit recovery is dependent on the candidate bit at the given site, $\tau_k=a$. Assuming no ergodicity breaking one can attain the variational part of the free energy density ((\ref{fd}) in the large $N$ limit) as
\begin{eqnarray}
f \! \!&=&\! \! \sum_a\! \int\! dh du W(a,h) \!{\hat W}(a,u)\log(1\!+\!\tanh(u)\tanh(h)) \nonumber\\
\!&+&\!\! \! \alpha \sum_a P_b(a) \left\{C \int du W(a,u) \log(\cosh u) \right.\nonumber  \\
\!&+&\!\! \! \left. \int \prod_{c=1}^{C} \left[du_c W(a,u_c)\right]\log\left(\cosh\left(\sum_{c=1}^C u_c\right)\right)\right\} \\
\!&+&\!\! \! \int \prod_{l=1}^L \left[dx_l d\phi(x_l) \sum_{a_l} dh_l W(a_l,h_l)\right] d\omega P_\nu(\omega)\log \mZ_I\nonumber\\
\mZ_I\!&=&\! \! \sum_\vtau \!\prod_{l=1}^L\! \left[\!\frac{\!\exp (h_l \tau_l)}{2\!\cosh(h_l)}\!\right]\!\exp\!\left\{\!-\mQ\left(\!\omega \!+\! \sum_{l=1}^L \!x_l \!a_l\!(1\!-\!\tau_l)\!\right)^2\!\right\}.\nonumber
\end{eqnarray}
Here $P_b$ is the true prior on transmitted bits, which we will assume to be uniform. We also assume the sparse ensemble with chip and user regularity for brevity. The distributions must be chosen to minimise the free energy, it is a near identical minimisation which gives rise to (\ref{saddle}). The pairs of field and bias distributions ${\hat W}$,$W$, in this case obey the saddlepoint equations
\begin{eqnarray}
  W(a,h) \!&\propto&\! \int \prod_{c=1}^{C-1} \left[du_c {\hat W}(a,u_c)\right] \delta\left(h - \sum_{c=1}^{C-1} u_c\right) \nonumber\\
  {\hat W}(a_L,u) \!&\propto&\!  \int \prod_{l=1}^{L-1} \left[ \phi(x_l) dx_l \sum_{a_l} W(a_l,h_l) dh_l\right] P_\nu(\omega)d\omega \nonumber\\
\!&\times&\! \delta\left(u - \sum_{\tau_L} \tau_L \log(\mZ(\tau_L))\right) \label{hatW_nomod}
\end{eqnarray}
Where $\mZ$ is the same quantity as (\ref{saddle}) upto the
substitution of $x_l$ by $a_l$. In this new case we have a
modified set of equations on distributions, as the dependence on
the root site cannot be factorised. Since we are considering
maximal rate both in the prior for sent message and inference
model we can argue by symmetry that $W(b,h)$ equals $W(-b,h)$.
This represents the intuitive statement that the probability of
correct reconstruction is independent of whether the sent bit is
$\pm 1$, however this is an ansatz rather than a result of the
calculation. The assumption can be tested by allowing convergence
restricted to the symmetric combination and testing small
perturbations in the antisymmetric part. A stronger test of the
ansatz is to allow the population dynamics to run with fully
independent distributions. To within numerical accuracy the
restricted solutions and those found in this larger space appear
to be consistent and the modulated and unmodulated sparse
ensembles become equivalent. At maximal rate the solution for the
unmodulated ensemble is information theoretically equivalent to
the unmodulated ensemble.

\section{Nature of the metastable solutions}
The exact results and numerical solutions (as indicated by example in Fig.~\ref{fren}) indicate several common features of the metastable state for both the sparse and dense systems. We investigate these points and present some simplified analysis of the energy landscape in this section. 
The results of the previous section provide insight into the probable nature of the state, and the fact that the sparse and dense systems are so similar qualitatively means that topology must play a relatively small role. The dynamical properties of the decoding algorithms reported for both cases appear to be an important common feature, while the sizes of solutions (as indicated by entropy) and bit error rates reduce the space of solutions to be considered.%
\subsection{Predictions for decoding failure in the marginal fields and couplings}
One can gain further insight by examining the interaction structure as a source of information, making analogies between other well studied disordered systems~\cite{Mezard:SGT}. The Hamiltonian may be re-written (upto constants) as
\begin{equation}
{\cal H}(\vtau) = - \left( \sum_{k \neq k'}  J_{k k'}\tau_k \tau_{k'} + \sum_k h_k \tau_k\right) \label{HamJh}
\end{equation}
which is a standard formulation in physics, where the set of couplings $J_{ij}$ and fields $h_i$ describe the problem
\begin{eqnarray}
J_{k,k'} \!\!\!&=&\!\!\! - \mQ \sum_\mu s_{\mu k} s_{\mu k'} \label{Jkk1}\\
h_k \!\!\!&=&\!\!\! 2 \mQ \sum_\mu y_\mu s_{\mu k} \label{hk1} = 2 \mQ \left[\sum_\mu s_{\mu k}^2\right] b_{k} \nonumber\\
\!\!\!&+&\!\!\! 2 \mQ \left(\left\{\sum_\mu  s_{\mu k} \sum_{k'(\neq k)} s_{\mu k'} b_{k'}\}\right\} + \left\{\sum_\mu \nu_\mu s_{\mu k}\}\right\}\right) \nonumber
\end{eqnarray}
Since the coupling term has no dependence on the sent bits $\vb$ the states induced by the couplings alone must be uncorrelated with the true solution. By contrast, the field term encodes a bias towards the sent vector combined with a pair of fields with no alignment along the correct solution (in expectation), but with some dependence thereof.

 The couplings and fields are strongly correlated through the code $\ms$. In the case of a dense code where $L\!\rightarrow\! K$ both marginal distributions over couplings and fields may be taken as Gaussian distributed through application of the central limit theorem with $N\!=\!K/\alpha$ large; the dense case gives
\begin{eqnarray}
P(J_{k,k'}) &=& {\cal N}\left(0,\frac{\mQ^2}{\alpha N}\right) \label{couplings}\;,\\
P(h_k) &=& {\cal N}\left(\frac{2 \mQ b_k}{\alpha}, \frac{(2\mQ)^2}{\alpha} + \frac{2 \mQ}{\alpha}\right) \;.\label{variance}
\end{eqnarray}
where ${\cal N}$ signifies the normal distribution. The first term of the field variance is negligable for the large system.

For the sparse code with BPSK one can instead note that the couplings are non-zero with probability ${L \choose 2}/{K \choose L}$ reflecting the enforced topology (\ref{sp}),(\ref{userreg}),(\ref{chipreg}), and in the non-zero cases take values $\pm \mQ/L$ with equal probability. In the field part one has a net positive field combined with two terms, the first term containing no noisy part gives a variance dependent on the site values and number of nearest neighbours (users connected through chips to user $k$), whereas the second is the sum of Gaussian random variables associated to each neighbouring chip. We approximate the distribution by a mean and variance to abbreviate this information, ignoring for convenience higher order moments as
\begin{equation}
 P(h_k) = {\cal N}\left( \frac{(2\mQ) b_k}{\alpha},  \frac{(L-1)(2\mQ)^2}{\alpha L} + \frac{2\mQ}{\alpha}\right) \label{Phkfields}\;.
\end{equation}
The $L\!-\!1$ prefactor is the average excess degree of the factor node in the chip regular ensemble (\ref{chipreg}), for the random graph ensemble (\ref{sp}) the value is $L$ (also with user regularity (\ref{userreg})). Using a non-regular code appears to impact upon the variance of the field but not the mean.

When one does not include the BPSK, the first two moments of the sparse distribution of local fields (\ref{Phkfields}) are unchanged but the couplings are entirely anti-ferromagnetic $\mQ/L$, again conforming to the underlying topology. At least for $\beta\!=\!1$ we have determined that the information theoretical quantities, and the population dynamics algorithm are equivalent for the two sparse ensembles considered. Therefore we expect only features common to the two models to be responsible for the metastability and other non-trivial properties in the large system limit.

We can now consider common features in the distributions.
In so far as a marginalised distribution might provide insight, it appears fairly clear that there is a competition between a mean dominated field producing good reconstruction and a variance dominated field leading to only marginal bias in favour of correct reconstruction. The field presumably projects into one of a number of local minima. When $\mQ$ is small the variance dominates and there is a weak net alignment with $\vb$. As one increases $\mQ$ the mean grows more quickly than the spread, so that in the large $\mQ$ limit the state is very orderly. By contrast as one increases $\alpha$ the mean is suppressed by comparison with the spread in the field (and in the couplings), so that one might expect the state to be variance dominated.

The marginal coupling distributions appear very different in the modulated models (sparse and dense) by comparison the unmodulated model. In the modulated model one has a random coupling, which one might expect would induce behaviour comparable to a random spin glass or an inverse of the Hopfield model~\cite{Mezard:SGT}, with a highly non-trivial distribution of local solutions (when ignoring the field). However, by investigation of the unmodulated model we see the space determined entirely by the couplings is in no way related to the modulation pattern, and hence the source of metastability cannot relate to this for our detector in the sparse case, the Hopfield analogy is certainly not useful. The second model is a random field Ising anti-ferromagnet, the former is a random field spin-glass, if the structure were a random graph with uncorrelated spin-spin edges (a Viana Bray model) we might expect behaviour to be quite comparable and described by a complicated energy landscape with many local minima -- in the absence of topological features the presence of metastability should not be a surprise, what is a surprise is that it appears for only a small range of parameters and has a bi-modal structure.

\subsection{Sources of metastability by analogy with CSPs}

What is important not to overlook in the above marginal link and field description is a consideration of the strong local correlations in graph topology, the interactions are formed in local cliques (fully connected sets of $L$ variables) and not independently. Although the fields are generated from an unusual ensemble they cannot be responsible for metastability, since in themselves they generate no long range correlations. We can first consider the role of couplings in the absence of a field. If one considers the details of the interaction structure one can observe that the ground state is closely related to random constraint satisfiability problems (CSP)~\cite{Mezard:SGT} such as the 'not all equal satisfiability' (NAE-SAT) model. Suppose chip connectivity of $L=3$ for all chips (hyperedges) in the system with an unmodulated sparse code, then the energy for the clique of dynamic variables (spins) attached to chip $\mu$ is $\sum b_k b_{k'}$ in the coupling part (\ref{HamJh}). This gives chip energy of either $3$, with all (modulated) spins equal, or $-1$ for any other assignments. The set of spin-assignment which simultaneously produces the fewest all equal cases (closest to the not all equal case satisfied case) are the ground state(s) of the system. The random NAESAT model is known to have a ground state set which is algorithmically non-trivial to find with variation of $\alpha$~\cite{ACIM01}. The fragmentation of the space (clustering) is understood to cause these features in many CSPs and statistical physics can produce exact descriptions of the correlations and other features of the thermodynamic solution. Figure \ref{fren} might be expected to reveal some corresponding phase transition in the underlying CSP with variation of $\alpha$. Thermodynamic features of the ground state correspond to properties of a maximum likelihood detector, which is closely related to the minimum bit error rate detector we analyse. 

Finally we must introduce the fields, afterall this is where the information about the transmitted bits exist. The field effectively define a vector in the energy landscape, and the energy must be minimised with respect to this direction (the energy landscape is effectively rotated). Using this analogy we can understand that the metastability arises out of the clustering of the underlying CSP reorientated by the field. One begins the search for the lowest energy in the vicinity of the matched-filter (field determined) solution, the local solution close to the encoded solution may be thermodynamically optimal but if the field projection is not into the cluster then local search methods are certain to fail. In solution spaces cf disjoint clusters one must work in a low noise regime, the field then almost certainly projects very close to the best state and local search is successful. This observation is consistent with the disappearance of suboptimal solutions at sufficiently high signal to noise ratios for all ensembles.

\section{Conclusion}
A comparison of the marginal coupling distributions in the two sparse cases indicates a substantial difference unlike a comparison between sparse and dense modulated code ensembles.
The quadratic Hamiltonian form seems to predict the appropriate regimes where decoding performance is weak by consideration of only the fields. 
The contrast between the two sparse ensembles suggests variance in the field is the most important factor in preventing successful decoding.
In one case the couplings are similar to those of a sparse spin glass, in the other the couplings are uniform, but anti-ferromagnetic.
When local topology is considered we see a connection to constraint satisfiability problems, which is a more convincing explanation of the origins of metastability.
To avoid metastable states in decoding we might hope to make use of the fact that we know the suboptimal states induced by the couplings are related to random CSPs, the ground states of which are for some parameterisations exactly solvable even on loopy graphs (with high probability), or have a well understood (asymptotic) state space structure. 
With a fragmented state space local search algorithms such as belief propagation may not converge, and other heuristic methods may be appropriate using a detailed knowledge of the CSP for example. 
It would also be interesting to further investigate what similarities exist between the modulated and unmodulated sparse codes in a wider range of detectors. The equivalence of modulated and unmodulated sparse codes in the minimum bit error rate detector should not apply to other detection methods or finite size systems, and hence in terms of practical performance of codes we may expect one ensemble to outperform the other.

\section*{Acknowledgment}
Support from EVERGROW, IP No.~1935 in FP6 of the EU and EPSRC grant
EP/E049516/1 are gratefully acknowledged.

\end{document}